# A stochastic programming approach for the scheduling of medical interpreting service under uncertainty

Abdulaziz Ahmed[a,b], Aida Jebali[c]

[a]Department of Health Services Administration, School of Health Professions, University of Alabama at Birmingham, Birmingham, AL 35233, USA

[b]Department of Biomedical Informatics and Data Science, Heersink School of Medicine, University of Alabama at Birmingham, Birmingham, Alabama 35233, USA

[c]SKEMA Business School, Université Côte d'Azur, France

**Abstract**

Limited English Proficiency (LEP) patients face an increased risk of adverse healthcare events due to communication barriers compared to English-speaking patients. These barriers can lead to misdiagnosis and prolonged hospital stays. The timeliness of medical interpreting services is indeed essential to improve the quality of care and health outcomes of LEP patients. However, given the hospital's budget restrictions, the efficiency of medical interpreting services should also be sought. This paper addresses the scheduling of medical interpreting services under uncertainty. The problem is formulated as a two-stage stochastic programming model that accounts for uncertainties related to the arrival of emergency patients and their service time, as well as the service time of outpatients. The model handles the hiring decisions of part-time interpreters and the assignment of full-time and hired part-time interpreters to LEP patients. The objective is to minimize the total cost, which encompasses full-time interpreters' overtime cost, the fixed and variable costs of part-time interpreters, and the penalty cost for not serving LEP patients on time. The model is first solved using the Sample Average Approximation (SAA) algorithm. Then, to overcome the computational burden of the SAA algorithm, the problem is solved using a well-designed Tabu Search (TS) algorithm. A real-life case study is used to validate the proposed solution algorithms and evaluate their performances. The results demonstrate the effectiveness of the proposed stochastic programming-based solutions in concurrently reducing the total cost and the waiting time compared to the Expected Value Problem (EVP) solution. The sensitivity analysis study further reveals how the increase of some key parameters, such as the fixed cost of part-time interpreters and the arrival rate of emergency patients with LEP, impact scheduling outcomes. Finally, some valuable managerial insights are provided for hospitals to better manage their medical interpreting services.

## 1. Introduction

According to the United States Census Bureau (2022), in 2019, there are nearly 68 million people (around 21% of the United States population) who speak languages other than English at their homes. In healthcare systems, patients, among those people, who have a limited ability to read, speak, write, or understand English are called Limited English Proficiency (LEP) patients. In 2019, there were approximately 25.6 million people with LEP in the United States (Ramirez et al. 2023). Markedly, the LEP patient population is one of the fastest-growing and most vulnerable segments of the society (Cohen et al. 2005). A recent study has evidenced that racial minority patients are generally admitted in lower-quality hospitals serving high proportions of LEP patients, even those who are living closer to higher-quality hospitals (Sliwinski et al. 2024). Because of their language deficiency, and the limited resources of the hospitals where they generally receive care, these patients are 20% more likely than English-speaking patients to experience longer length of stays (LOSs), higher readmission rates, and more adverse healthcare incidents, some of which could lead to death (Divi et al. 2007). Conspicuously, overcoming language barriers by providing adequate medical interpreting services would contribute not only to curbing these suboptimal care outcomes (Beagley et al. 2020) but also to reducing segregation in hospital care.

Medical interpreting service is crucial for delivering safe and efficient healthcare services for LEP patients. It is indeed essential to ensure effective communication between LEP patients and healthcare providers. Evidently, ineffective communication between providers and LEP patients can undermine trust in the quality of delivered medical services and decrease the likelihood of a patient receiving proper follow-up (Flores et al. 2002). Language barriers lead to misunderstanding diagnoses, medical test results, treatment plans, and disparities in prescriptions. In 2010, the School of Public Health and National Health Law Program at the University of California Berkeley found that, of 1373 malpractice claims, 35 cases were caused by inadequate interpreting services. Those cases included brain damage, dismemberment, death, and other severe medical condition (Quan and Lynch 2010). Moreover, as underlined in (Jacobs et al. 2004), the availability of interpreting services fosters the receipt of preventive care by LEP patients, thereby helping prevent health complications. Interpreters also help physicians to understand symptoms, and so diagnosing an LEP patient's condition properly, which allows to develop effective treatment plans and results in a more efficient use of hospital resources (Ku and Flores 2005).

Providing interpreting services is a legal requirement for hospitals (Jacobs et al. 2018), and failure to do so may expose them to legal consequences. According to Title VI of the Civil Rights Act of 1964 (U.S. Department of Health and Human Services, 2024), denying or delaying medical services to LEP patients because of language barriers is even considered a form of discrimination. In the same vein, the Joint Commission, makes the availability of interpreting services a requirement for hospital accreditation (Wilson-Stronks and Galvez 2007). Interpreting service constitutes a financial burden on hospitals, given that private payers do not reimburse hospitals for it, and Medicaid covers only about 10% of the total cost (Wilson 2013). At the same time, providing such service not only enhances the accessibility and the quality of healthcare services delivered to LEP patients but also improves the efficiency of hospital resource utilization.

In this paper, we develop a stochastic programming approach for scheduling medical interpreting service. We conducted the study based on real data collected from a partner hospital located near the metropolitan area of Minneapolis/St. Paul in Minnesota in the United States. According to the American Immigration Council (2020), about 10% of Minnesota residents are immigrants, and most of them live near Minneapolis/St. Paul area. Our partner hospital is mandated to provide interpreting services for LEP patients in inpatient, outpatient, and emergency units. It is very important for hospital managers to investigate the related scheduling problem to ensure timely and cost-effective service to LEP patients. The hospital employs full-time interpreters, and whenever needed, part-time interpreters are hired to fulfill the demand. The objective of scheduling is to provide the most efficient interpreting service to LEP inpatients, outpatients, and emergency patients while minimizing their waiting times. This allows the interpreting service to thrive with service quality, efficiency, and patient satisfaction. Scheduling interpreting services is challenging due to the inherent sources of uncertainty. Two main types of uncertainties should be accounted for: (1) those stemming from the arrival process of LEP patients to the emergency department (ED) and (2) those tied to service duration. Clearly, the unavailability of an interpreter at the arrival time of an LEP patient to the ED might increase the waiting time of the patient and so could lead to a prolonged LOS of the patient in the hospital (Gurazada et al. 2022) and in some extreme cases to a life-threatening situation. Furthermore, when an inpatient or an outpatient does not show up, the part-time interpreter gets paid even if she does not provide the service, which is a cost-draining situation for a hospital. Another aspect that makes interpreting service scheduling more difficult is the many languages spoken in the United States, given that matching an LEP patient's language and the language an interpreter can work with is a must.

To tackle the problem, we develop a two-stage stochastic programming model for scheduling medical interpreters while considering the uncertainty related to service duration and the arrival of emergency patients with LEP, including their requirements in terms of interpreting service (i.e., language). First, a featured Sample Average Approximation (SAA) algorithm is used to solve the problem. Then, a simulation-based optimization approach using tabu search is devised to solve the problem in a more reasonable time and so overcome the computational burden of the SAA approach. The novelty of this study stems from the following: (1) up to our knowledge, this is the first work in the literature that studies interpreter scheduling problems under uncertainty; (2) we develop a two-stage programming model to schedule interpreters while explicitly taking into account the sources of uncertainty mentioned above; (3) we consider all the dimensions of the service including interpreters' availability, language match, etc.

The remainder of this paper is organized as follows. Section 2 reviews the most relevant works on interpreting services, and staffing and scheduling problems. Section 3 presents the addressed problem and the proposed formulation. Section 4 exposes the developed solution methods. Section 5 reports computational experiments and results. Finally, section 6 concludes this work with important findings and recommendations for future research.

## 2. Literature review
### 2.1 Importance of medical interpreting services

Many studies have been devoted to investigating the impact of medical interpreting services on the quality of care provided to LEP patients. For example, Estrada and Messias (2017) studied how language asymmetry between patients and providers impacts the quality of healthcare services. The study was conducted in the Charlotte metropolitan area of North Carolina. It was found that the communication process between LEP patients and healthcare providers affects providers' decisions and LEP patients' understanding and potential compliance. More recently, Torresdey et al. (2024) found that increased time with medical interpreters improves communication between clinicians and patients with LEP. Interpreting service mitigates the impact of language asymmetry. Many researchers also observed that when medical interpreting services are not provided, LEP patients are more likely to stay longer in a hospital and have a higher readmission rate. Lindholm et al. (2012) investigated the role of language services with respect to LEP patients' LOS and hospital readmission. They analyzed the admission data for LEP patients over three years (between May 1, 2004 and April 30, 2007). Then, they calculated the LOS for all patients and the readmission rate. They found that LOS was significantly longer when no interpreting service was provided at admission and/or discharge. John-Baptiste et al. (2004) demonstrated that LEP patients stay 6% more than English-speaking patients while Rawal et al. (2019) showed that LEP patients with chronic obstructive pulmonary disease and heart failure are more likely to be readmitted or return to a hospital than English-speaking patients. Recently, Kwan et al. (2023) conducted an integrative review to study the impact of medical interpreter services on hospital care outcomes. They analyzed 37 articles, highlighting how language services improve communication quality, patient safety, and satisfaction while potentially reducing hospital costs.

Furthermore, the absence of interpreting services contributes to a misunderstanding of LEP patients' conditions and medication. To study the impact of language barriers on understanding medical information,

Wilson et al. (2005) compared LEP patient visits with and without the presence of a physician who speaks the same language. They found that when there is language asymmetry between an LEP patient and the physician, the patient does not fully understand the medical situation, medication use, and medication side effects. In addition, language barriers affect the quality of some medical tests. Moreover, Taffel et al. (2020) found that the quality of MRI images was worse when no interpreting service was provided for LEP patients.

Therefore, LEP patients are more likely to experience medical errors. Divi et al. (2007) analyzed more than 1000 adverse events collected for LEP patients and English-speaking patients. It was found that 49.1% of the incidences happened to LEP patients versus 29.5% of them happened to English-speaking patients. More importantly, 46.8% of adverse events affecting LEP patients range from moderate to fatal, compared to 24.4% for English-speaking patients. 52.4% of incidences that occurred to LEP patients were indeed caused by miscommunication. In the era of pandemics (e.g., COVID-19), interpreting services helped save LEP patients' lives. When an LEP patient does not have enough information about the protective measures from such a virus, avoiding infection becomes more difficult. Also, LEP patients may have difficulties understanding the symptoms of COVID-19, and how to access a healthcare service when needed. Understanding doctors and nurses is difficult without proper access to interpreting services (Simon 2020). As noted, the studies presented so far focus on the positive impact of interpreting services on the health outcomes of LEP patients. However, none of these studies proposed an advanced modeling approach for the interpreter staffing and scheduling problem, even though this could significantly contribute to improving the level and the quality of interpreting service in the hospital and so LEP patient health outcomes.

## 2.2  Scheduling of medical interpreting service

The scheduling of medical interpreting service involves the staffing and the scheduling of medical interpreters. Specifically, it intends to determine the staffing requirements for interpreters based on LEP patient demand per language, while efficiently assigning them to interpreting tasks. This assignment considers factors such as interpreter skills (languages), availability, task duration, and release time. The goal in this workforce staffing and scheduling problem is to optimize the utilization of interpreters to timely meet service demands, ensure quality, and contain costs.

Workforce scheduling has been studied extensively across different industries, including healthcare, technical services, and security. Castillo-Salazar et al. (2016) conducted a literature review on workforce scheduling and routing problems. They summarized the key characteristics of the problem, which are time windows, transportation, required skills, service time, connected activities, teaming, and clustering of tasks. Although the interpreter scheduling problem has similarities with many workforce scheduling problems addressed in the literature, it differs from them in various aspects, such as the higher level of inherent uncertainty. For example, Çakırgil et al. (2020) developed a model to optimize the daily scheduling and routing of maintenance multi-skill workforce for an energy distribution company. They mentioned that around 90% of the maintenance requests are known in advance. Henceforth, the level of uncertainty pertaining to the interpreter scheduling problem is relatively higher, given the large number of stochastic parameters it encompasses. These uncertainties are tied to the arrival of emergency patients with LEP, their service time and the service time of LEP outpatients. Notably, emergency patients with LEP could represent a large proportion of the served LEP patients.

As far as the healthcare sector is concerned, there is an abundant literature on nurse and physician scheduling problems. In nurse scheduling, both preference satisfaction and cost must be considered. Nurse scheduling is usually accomplished in two stages. Firstly, a schedule is prepared to meet the collective agreement for the regular working time, which helps to manage personnel shortages and surpluses. Then, any remaining shortages are addressed by using overtime or on-call nurses (Bruni and Detti 2014). In physicians scheduling, preference satisfaction is often the primary driving force of scheduling, as retaining physicians is a common challenge for hospital administrations (Bruni and Detti 2014). For certain specialties, the scheduling can consider both physician on regular duty and on-call (Rath and Rajaram 2022). Similarly to medical interpreter scheduling, each day the hospital manager should decide on the number of on-call (part-time) physicians needed for the following. Total cost minimization is generally the primary objective in daily scheduling.

The main constraints in interpreter scheduling process pertain to language match between interpreters and LEP patients, full-time and part-time interpreter availability, and LEP patient categories (e.g., inpatient, outpatient, and emergency patient). Furthermore, this problem can be considered as a multi-skilling workforce staffing and scheduling problem given that some interpreters are cross-trained to ensure service for several languages (Çakırgil et al. 2020). Although there are several benefits to multi-skilling in service Easton (2014), the recent literature review conducted by (Afshar-Nadjafi 2021) shows that it has seldom been considered in hospital staff scheduling. (Othman et al. 2015) is the only paper that addressed the scheduling of multi-skilled medical staff (including, nurses, care assistants, and pediatricians) to perform different tasks in a pediatric ED. The objective is to minimize patients' waiting time, the workload of the medical staff as well as the response time.

Various approaches have been proposed to deal with workforce staffing and scheduling problems. Such approaches involve mathematical modeling, simulation, and metaheuristics. Abdalkareem et al. (2021) conducted a review study on scheduling problems in healthcare. They demonstrated that most of the approaches used to solve scheduling problems are based on metaheuristics. Moreover, Afshar-Nadjafi (2021) showed that 61.8% of the approaches used to solve multi-skilling scheduling problems were heuristic-based approaches (heuristics, metaheuristics and matheuristics). 27.5% of the problems were solved by exact algorithms such as the branch and bound of off-the-shelf solvers, 6.9% were solved using simulation, and 3.8% of the problems were not solved. When it comes to modeling, most studies have used Mixed Integer Programming (MIP) to formulate the scheduling problem they addressed (Afshar-Nadjafi 2021). Two-stage stochastic programming (Easton 2011, Bagheri et al. 2016, Ta et al. 2021), simulation (Angelidis et al. 2013), system dynamic (Alvanchi et al. 2012), heuristics (Fu et al. 2024), and hybrid approaches (Avramidis et al. 2010) were utilized to handle stochastic environment. For example, Bagheri et al. (2016) used two-stage stochastic programming to model a nurse scheduling problem and solved it using the SAA approach.

Very few studies have been conducted to investigate the interpreter scheduling problem. This has been concluded after checking different search engines, such as Google Scholar and Scopus, and trying various keywords, such as interpreter scheduling, translator scheduling. Chiam et al. (2017) proposed a simulation-optimization model for finding the break-even point for hiring full-time and part-time Spanish and Mandarin interpreters at Christiana Care Health System. However, their model was simple and considered only a limited number of languages. Ahmed and Hamasha (2018) developed an MIP model for interpreter scheduling. However, only LEP outpatients were considered, and it was assumed that all of them had the same service time. Ahmed and Frohn (2021) developed the most advanced model for scheduling medical interpreters. They developed an MIP programming model for assigning full-time and part-time interpreters and then solved the model using the Pulp package. However, they did not consider the stochastic aspects of the problem (e.g., service time and emergency patient arrival). Also, they assumed that LEP patients get the service on time. Therefore, this is, to the best of our knowledge, the first work that studies interpreter scheduling problem under uncertainty and taking into account different categories of patients with LEP.

## 3. Problem description and model formulation

### 3.1 Problem description

Interpreting service should be provided for three categories of LEP patients: (1) inpatients, (2) outpatients, and (3) emergency patients, by a set of full-time and part-time interpreters. The scheduling of medical interpreting service is subject to two main types of uncertainties: (1) those stemming from the arrival process of LEP patients to the ED and (2) those tied to service duration. The decision on the hiring of part-time interpreters should be confirmed to the contracted medical interpreting service agency at the end of each day for the next day. Therefore, the schedule of medical interpreting service, including the staffing of part-time interpreters, is built at the evening of each day for the next day. This process is justified, first, by the fact that the number of LEP inpatients and scheduled LEP outpatients are only unveiled the evening before the targeted day. In addition, this allows to confirm the hiring of part-time interpreters based on a more accurate forecast of the arrival rates of emergency patients with LEP and of the mean and standard deviation of service duration. Indeed, as revealed in (Ta et al. 2021), and especially in the case of emergency services, the uncertainty is generally reduced when we get closer to the scheduling period. The objective is to achieve the best trade-off between the cost and the quality of service (QoS) provided to LEP patients. The QoS is improved by reducing LEP patients' waiting time for interpreting services.

A two-stage stochastic programming model is devised for the considered medical interpreting service scheduling problem. The hiring decisions of part-time interpreters, made in the evening, just before the targeted day, are the first-stage decisions. The scheduling decisions are taken once the uncertainty has been resolved and as such, are second-stage decisions. Scheduling interpreting services is challenging due to the several inherent sources of uncertainty. Recall that two main types of uncertainties should be accounted for, (1) those stemming from the arrival process of LEP patients to the ED and (2) those tied to service duration. At this level, it is noteworthy that the uncertainty related to the arrival of emergency patients involves two dimensions: (1) the uncertainty in the arrival time and (2) the uncertainty in terms of interpreting service (i.e., language). Furthermore, the uncertainty in duration pertains to both the service provided to emergency patients and outpatients. As far as LEP inpatients are concerned, the duration of service is ensured for the entire period and so is known.

### 3.2 Assumptions

- Unlike emergency patients, the arrival of inpatients and outpatients is scheduled in advance and, therefore, known.
- Each LEP patient should be assigned to an interpreter with the needed interpreting skill (language).
- The duration of an interpreting session is deterministically known for inpatients, while it is uncertain for outpatients and emergency patients.
- The service provided by part-time interpreters incurs both fixed and variable costs. The fixed cost includes service for a certain time threshold. Beyond this time threshold, the pay of a part-time interpreter involves a variable cost.

### 3.3 Problem formulation

Based on these assumptions, the problem is formulated as a two-stage stochastic programming model. The notation used in the model formulation is given below.

Sets and indices

| | |
|---|---|
| $\bar{I}$ | Set of full-time interpreters indexed by $i$, $i = 1, 2, \dots, I$ |
| $\bar{J}$ | Set of part-time interpreters indexed by $j$, $j = 1, 2, \dots, J$ |
| $\bar{T}$ | Set of time periods indexed by $t$, $t = 1, 2, \dots, T$ |
| $\Xi$ | Set of scenarios indexed by $\xi$, $\xi = 1.. |\Xi|$ |
| $\bar{N}1$ | Set of LEP inpatients indexed by $n$, $n = 1, 2, \dots, N1$ |
| $\bar{N}2$ | Set of LEP outpatients indexed by $n$, $n = N1 + 1, N1 + 2, \dots, N1 + N2$ |
| $\bar{N}3_\xi$ | Set of LEP emergency patients indexed by $n$, $n = N1 + N2 + 1, \dots, N_\xi = N1 + N2 + N3_\xi$ |
| $\bar{N}_\xi$ | Set of LEP patients indexed by $n$, $n = 1, 2, \dots, N_\xi$ |

Parameters

| | |
|---|---|
| $OC_i^f$ | Unit overtime cost for full-time interpreter $i$ |
| $F_j^p$ | Fixed cost of part-time interpreter $j$ |
| $WF_j^p$ | Service time of part-time interpreter $j$ included in the fixed cost |
| $C_j^p$ | Unit pay rate of part-time interpreter $j$ for the service time ensured beyond the one included in the fixed cost |
| $P_n$ | Unit penalty cost for waiting for patient $n$, $n = 1, 2, \dots, N_s$ |
| $z_{in}^f$ | =1 if full-time interpreter $i$ can serve patient $n$ ($n = 1, 2, \dots, N1 + N2$), 0 otherwise |
| $z_{jn}^p$ | =1 if part-time interpreter $j$ can serve patient $n$ ($n = 1, 2, \dots, N1 + N2$), 0 otherwise |
| $z3_{in\xi}^f$ | =1 if, under scenario $\xi$, full-time interpreter $i$ can serve patient $n$ ($n = N1 + N2 + 1, \dots, N_s$), 0 otherwise |
| $z3_{jn\xi}^p$ | =1 if, under scenario $\xi$, part-time interpreter $j$ can serve patient $n$ $(n = N1 + N2 + 1, \dots, N_\xi)$, 0 otherwise |
| $A_{it}^f$ | =1 if full-time interpreter $i$ is available to provide service at time period $t$, 0 otherwise |
| $A_{jt}^p$ | =1 if part-time interpreter $j$ is available to provide service at time period $t$, 0 otherwise |
| $\lambda 1_{nt}$ | =1 if patient $n$, $n \in \bar{N}1$, arrives at time period $t$, 0 otherwise |
| $t1_n$ | Time period over which patient $n$, $n \in \bar{N}1$, arrives ($t1_n = \sum_{t=1}^{T} t\lambda 1_{nt}$) |
| $\lambda 2_{nt}$ | =1 if patient $n$, $n \in \bar{N}2$, arrives at time period $t$, 0 otherwise |
| $t2_n$ | Time period over which patient $n$, $n \in \bar{N}2$, arrives ($t2_n = \sum_{t=1}^{T} t\lambda 2_{nt}$) |
| $\lambda 3_{nt\xi}$ | =1 if, under scenario $\xi$, patient $n$, $n \in \bar{N}3_s$, arrives at time period $t$, 0 otherwise |
| $t3_{\xi n}$ | Time period, under scenario $\xi$, over which patient $n$, $n \in \bar{N}3_s$, arrives ($t3_{\xi n} = \sum_{t=1}^{T} t\lambda 3_{nt\xi}$) |
| $D1_n$ | Duration of interpreting session for patient $n$, $n \in \bar{N}1$ |
| $D2_{n\xi}$ | Duration of interpreting session for patient $n$, $n \in \bar{N}2$, under scenario $\xi$ |
| $D3_{n\xi}$ | Duration of interpreting session for patient $n$, $n \in \bar{N}3_\xi$, under scenario $\xi$ |

$p_\xi$     Probability of scenario $\xi$

$WR_i^f$     Regular working time of full-time interpreter $i$

$\varepsilon$     A constant with a small positive value (it can be given a value less than or equal to $1/T.|\Xi|.N$ with $N = \max_{\xi=1..|\Xi|} N_\xi$)

$\alpha$     A constant greater than 1

Decision variables

First-stage variables

$w_j$     =1 if part-time interpreter $j$ is hired, 0 otherwise

Second-stage variables

$x_{int\xi}$     =1 if full-time interpreter $i$ starts interpreting session for patient $n$ $(n \in \overline{N}_\xi)$ at time period $t$ under scenario $\xi$, 0 otherwise

$y_{jnt\xi}$     =1 if part-time interpreter $j$ starts interpreting session for patient $n$ $(n \in \overline{N}_\xi)$ at time period $t$ under scenario $\xi$, 0 otherwise

$H_{i\xi}^f$     Total interpreting service time for full-time interpreter $i$ under scenario $\xi$

$H_{j\xi}^p$     Total interpreting service time for part-time interpreter $j$ under scenario $\xi$

$U_{n\xi}$     Waiting time of patient $n$ $(n \in \overline{N}_\xi)$ under scenario $\xi$

$ST_{j\xi}$     Service time ensured by part-time interpreter $j$ beyond the one included in the fixed cost under scenario $\xi$

$OT_{i\xi}$     Overtime of full-interpreter $i$ under scenario $\xi$

The proposed two-stage stochastic programming model is formulated as follows:

$$\text{Min} \sum_{j=1}^{J} F_j^p w_j + \sum_{\xi=1}^{|\Xi|} p_\xi \left( \sum_{j=1}^{J} C_j^p ST_{j\xi} + \sum_{i=1}^{I} OC_i^f OT_{i\xi} + \sum_{n=1}^{N_\xi} P_n U_{n\xi} \right) \quad (1)$$

Subject to:

$$\sum_{i=1}^{I} \sum_{t=1}^{T} x_{int\xi} + \sum_{j=1}^{J} \sum_{t=1}^{T} y_{jnt\xi} \leq 1 \quad \forall n \in \overline{N}_\xi, \xi \in \Xi \quad (2)$$

$$\sum_{n=1}^{N_\xi} x_{int\xi} \leq A_{it}^f \quad \forall i \in \overline{I}, t \in \overline{T}, \xi \in \Xi \quad (3)$$

$$\sum_{n=1}^{N_\xi} y_{jnt\xi} \leq A_{jt}^p \quad \forall j \in \overline{J}, t \in \overline{T}, \xi \in \Xi \quad (4)$$

$$x_{in'h\xi} \leq (1 - x_{int\xi}) \ \forall \ i \in \bar{I}, t \in \bar{T}, \xi \in \Xi, \ n \in \bar{N}1, n' \in \bar{N}_\xi \backslash \{n\}, h \in \{t + 1, \ldots, \min(T, t + D1_n - 1)\} \quad (5)$$

$$x_{in'h\xi} \leq (1 - x_{int\xi}) \ \forall \ i \in \bar{I}, t \in \bar{T}, \xi \in \Xi, \ n \in \bar{N}2, n' \in \bar{N}_\xi \backslash \{n\}, h \in \{t + 1, \ldots, \min(T, t + D2_{n\xi} - 1)\} \quad (6)$$

$$x_{in'h\xi} \leq (1 - x_{int\xi}) \ \forall \ i \in \bar{I}, t \in \bar{T}, \xi \in \Xi, n \in \bar{N}3_\xi, n' \in \bar{N}_\xi \backslash \{n\}, h \in \{t + 1, \ldots, \min(T, t + D3_{n\xi} - 1)\} \quad (7)$$

$$y_{jn'hs} \leq (1 - y_{jnt\xi}) \forall j \in \bar{J}, t \in \bar{T}, \xi \in \Xi, n \in \bar{N}1, n' \in \bar{N}_\xi \backslash \{n\}, h \in \{t + 1, \ldots, \min(T, t + D1_n - 1)\} \quad (8)$$

$$y_{jn'h\xi} \leq (1 - y_{jnt\xi}) \forall i \in \bar{J}, t \in \bar{T}, \xi \in \Xi, n \in \bar{N}2, n' \in \bar{N}_\xi \backslash \{n\}, h \in \{t + 1, \ldots, \min(T, t + D2_{n\xi} - 1)\} \quad (9)$$

$$y_{jn'h\xi} \leq (1 - y_{jnt\xi}) \forall i \in \bar{J}, t \in \bar{T}, \xi \in \Xi, n \in \bar{N}3_\xi, n' \in \bar{N}_\xi \backslash \{n\}, h \in \{t + 1, \ldots, \min(T, t + D3_{n\xi} - 1)\} \quad (10)$$

$$\sum_{i=1}^{I} \sum_{t=1}^{T} t x_{int\xi} + \sum_{j=1}^{J} \sum_{t=1}^{T} t y_{jnt\xi} + \alpha T (1 - \sum_{i=1}^{I} \sum_{t=1}^{T} x_{int\xi} - \sum_{j=1}^{J} \sum_{t=1}^{T} y_{jnt\xi}) - t1_n \leq U_{n\xi} \ \forall \xi \in \Xi, n \in \bar{N}1 \quad (11)$$

$$\sum_{i=1}^{I} \sum_{t=1}^{T} t x_{int\xi} + \sum_{j=1}^{J} \sum_{t=1}^{T} t y_{jnt\xi} + \alpha T (1 - \sum_{i=1}^{I} \sum_{t=1}^{T} x_{int\xi} - \sum_{j=1}^{J} \sum_{t=1}^{T} y_{jnt\xi}) - t2_n \leq U_{n\xi} \ \forall \xi \in \Xi, n \in \bar{N}2 \quad (12)$$

$$\sum_{i=1}^{I} \sum_{t=1}^{T} t x_{int\xi} + \sum_{j=1}^{J} \sum_{t=1}^{T} t y_{jnt\xi} + \alpha T (1 - \sum_{i=1}^{I} \sum_{t=1}^{T} x_{int\xi} - \sum_{j=1}^{J} \sum_{t=1}^{T} y_{jnt\xi}) - t3_{\xi n} \leq U_{n\xi} \ \forall \xi \in \Xi, n \in \bar{N}3_\xi \quad (13)$$

$$x_{int\xi} = 0 \ \forall \ i \in \bar{I}, n \in \bar{N}1, t < t1_n, \xi \in \Xi \quad (14)$$

$$y_{jnt\xi} = 0 \ \forall \ j \in \bar{J}, n \in \bar{N}1, t < t1_n, \xi \in \Xi \quad (15)$$

$$x_{int\xi} = 0 \ \forall \ i \in \bar{I}, n \in \bar{N}2, t < t2_n, \xi \in \Xi \quad (16)$$

$$y_{jnt\xi} = 0 \ \forall \ j \in \bar{J}, n \in \bar{N}2, t < t2_n, \xi \in \Xi \quad (17)$$

$$x_{int\xi} = 0 \ \forall \ i \in \bar{I}, n \in \bar{N}3_\xi, t < t3_{\xi n}, \xi \in \Xi \quad (18)$$

$$y_{int\xi} = 0 \ \forall j \in \bar{J}, n \in \bar{N}3_\xi, t < t3_{\xi n}, \xi \in \Xi \quad (19)$$

$$x_{int\xi} \leq z_{in}^f \ \forall \ i \in \bar{I}, n \in \bar{N}1 \cup \bar{N}2, t \in \bar{T}, \xi \in \Xi \quad (20)$$

$$y_{jnt\xi} \leq z_{jn}^p \ \forall \ j \in \bar{J}, n \in \bar{N}1 \cup \bar{N}2, t \in \bar{T}, \xi \in \Xi \quad (21)$$

$$x_{int\xi} \leq z3_{in\xi}^f \ \forall \ i \in \bar{I}, n \in \bar{N}3_\xi, t \in \bar{T}, \xi \in \Xi \quad (22)$$

$$y_{jnt\xi} \leq z_{jn\xi}^p \ \forall j \in \bar{J}, n \in \bar{N}3_\xi, t \in \bar{T}, \xi \in \Xi \quad (23)$$

$$\sum_{n=1}^{N1}\sum_{t=1}^{T} x_{int\xi} D1_n + \sum_{n=N1+1}^{N1+N2}\sum_{t=1}^{T} x_{int\xi} D2_{\xi n} + \sum_{n=N1+N2+1}^{Ns}\sum_{t=1}^{T} x_{int\xi} D3_{\xi n} = H_{i\xi}^{f} \quad \forall\, i \in \bar{I}, \xi \in \Xi \tag{24}$$

$$\sum_{n=1}^{N1}\sum_{t=1}^{T} y_{jnt\xi} D1_n + \sum_{n=N1+1}^{N1+N2}\sum_{t=1}^{T} y_{jnt\xi} D2_{\xi n} + \sum_{n=N1+N2+1}^{N\xi}\sum_{t=1}^{T} y_{jnt\xi} D3_{\xi n} = H_{j\xi}^{p} \quad \forall\, j \in \bar{J}, \xi \in \Xi \tag{25}$$

$$w_j \geq \varepsilon \sum_{\xi=1}^{S}\sum_{n=1}^{N_s}\sum_{t=1}^{T} y_{jnt\xi} \quad \forall j \in \bar{J} \tag{26}$$

$$H_{i\xi}^{f} - WR_i^{f} \leq OT_{i\xi} \quad \forall\, i \in \bar{I}, \xi \in \Xi \tag{27}$$

$$H_{j\xi}^{p} - WF_j^{p} \leq ST_{j\xi} \quad \forall j \in \bar{J}, \xi \in \Xi \tag{28}$$

$$w_j, x_{int\xi}, y_{jnt\xi} \in \{0,1\} \quad \forall j \in \bar{J}\, \forall\, i \in \bar{I}, n \in \bar{N}_\xi, t \in \bar{T}, \xi \in \Xi \tag{29}$$

$$U_{n\xi}, H_{i\xi}^{f}, H_{j\xi}^{p}, OT_{i\xi}, ST_{j\xi} \geq 0, \forall\, n \in \bar{N}_\xi, \forall\, i \in \bar{I}, \forall\, j \in \bar{J}, \xi \in \Xi \tag{30}$$

The objective function (1) aims to minimize the total cost, composed of the first-stage fixed cost of hiring part-time interpreters, and the expected value of the second-stage cost that includes the service cost of part-time interpreters beyond the one included in the fixed cost, the overtime cost of full-time interpreters, and the penalty cost incurred due to LEP patients' waiting time for service. Constraints (2) guarantee that, under each scenario, at most a full-time or a part-time interpreter is assigned to an LEP patient. Constraints (3) and (4) ensure that, for each scenario, a full-time or a part-time interpreter can initiate at most one interpreting session at the beginning of a time period, during which she is available to provide service. Constraints (5)-(7) enforce the continuity of interpreting sessions given by full-time interpreters to LEP inpatients, outpatients and emergency patients, respectively. Similarly, constraints (8)-(10) guarantee the continuity of interpreting sessions ensured by part-time interpreters. As such, constraints (5)-(10) imply that once an interpreting session starts, it cannot be interrupted. The combination of constraints (11)-(13) with the objective function allows us to calculate the waiting time for interpreting service for LEP inpatients, outpatients, and emergency patients, respectively. As can be noted, if under a certain scenario, an LEP patient cannot be assigned to an interpreter, then their waiting time takes on the value of "$\alpha T$ – their arrival time" with $\alpha > 1$. Therefore, the penalty cost induced by the waiting time increases with $\alpha$. Constraints (14)-(19) ensure that an interpreting session can only start after the arrival of the LEP patient. Constraints (20)-(23) guarantee the assignment of each LEP patient to an interpreter with the required interpreting skills. Constraints (24) and (25) determine the total service time for full-time and part-time interpreters, respectively. The combination of constraints (26) with the objective function ensures that a part-time interpreter is hired if they are assigned to at least one interpreting session. The combination of constraints (27) with the objective function allows to determine the overtime for each full-time interpreter, while the combination of constraints (28) with the objective function allows to determine the service time provided by each part-time interpreter beyond the one included in the fixed cost. Constraints (29) and (30) are domain constraints.

## 4. Solution methodology

### 4.1 SAA approach

#### 4.1.1 SAA formulation

A featured sample average approximation (SAA) algorithm is used to solve the formulated two-stage stochastic programming model (Shapiro et al. 2002). For that, we start by generating $S$ independent scenarios ($S<|\Xi|$) using Monte Carlo simulation. Each generated scenario $s$ is indeed a possible realization of the following independent random variables: the arrival time of emergency patients with LEP for each needed interpreting skill and the interpreting session duration for each LEP outpatient and emergency patient. By using these $S$ scenarios (also referred to as a sample of size $S$), the stochastic model is approximated by a deterministic one, referred to as the SAA model. To obtain the SAA model, we replace the set of scenarios $\Xi$ and their indexes $\xi$ in the two-stage stochastic programming model by the set of generated scenarios $S$ and the corresponding indexes $s$. As far as the objective function is concerned, the expected value of the second-stage cost is approximated by the sample average over the generated $S$ scenarios. For the sake of simplification, the probability of each scenario $s$ is approximated by $1/S$. Then, the SAA model is solved to optimality to find a good approximate solution to the two-stage stochastic programming model. It is worth noting that the quality of the optimal solution of the SAA model improves when the sample size $S$ increases; it even converges to an optimal solution of the two-stage stochastic programming model with probability approaching one exponentially fast when $S$ tends to infinity (Shapiro and Homem-de-Mello 2000). However, the SAA model incorporating samples of large size is computationally intractable. Therefore, selecting the sample size $S$ requires finding a balance between the quality of an SAA solution and the computational time needed to obtain it. To overcome the expected computational burden of the SAA model, we start by simplifying it while leveraging some peculiar features of the studied problem and then strengthening the updated formulation by adding some valid cuts.

#### 4.1.2 Simplification of the SAA model

According to the practice of our partner hospital, but also in many other hospitals, an interpreter with adequate skills is reserved for each elective inpatient with LEP several days before their admission. Non-elective inpatients will continue with the interpreter who served them at the ED, prior to their admission. Moreover, an interpreter is pre-assigned to an inpatient for the entire day she is staying in the hospital. In this context, the inpatients and the interpreters pre-assigned to them can be excluded from the two-stage stochastic programming model and its sampling-based deterministic equivalent, i.e., the SAA model. This allows to simplify the SAA model and reduce its size in the number of decision variables and constraints and so its complexity. In this simplified version of the SAA model, $\bar{N}_s$ (with cardinal $N_s$) is reduced to $\bar{N}'_s$ (with cardinal $N'_s$) by excluding from it the set $\bar{N}1$. In addition, the sets $\bar{I}$ and $\bar{J}$ (with cardinal $I$ and $J$, respectively) are reduced to $\bar{I}'$ and $\bar{J}'$ (with cardinal $I'$ and $J'$, respectively) by excluding the interpreters pre-assigned to inpatients. Evidently, in this new version of the SAA model, $\bar{N}_s$, $N_s$, $\bar{I}$, $I$, $\bar{J}$ and $J$ in the objective function and the remaining constraints, are replaced by $\bar{N}'_s$, $N'_s$, $\bar{I}'$, $I'$, $\bar{J}'$ and $J'$, respectively. In addition, the set $\bar{N}2$ is thereafter indexed by $n=1..N2$ while $\bar{N}3_s$ is indexed by $n=N2+1.. N'_s = N2 + N3_s$.

### 4.1.3 Strengthening of the SAA model

Valid inequalities (31) and (32) are thereafter introduced to strengthen the SAA model formulation. The objective of adding valid cuts is to solve the model faster by eliminating regions that do not contain feasible solutions (Cornuéjols 2008, Tlahig et al. 2013, Diabat and Jebali 2021).

$$w_j \leq \sum_{s=1}^{S} \sum_{n=1}^{N'_s} \sum_{t=1}^{T} y_{jnts} \quad \forall j \in \bar{J}' \tag{31}$$

$$w_j \geq y_{jnts} \quad \forall j \in \bar{J}', s \in \{1..S\}, n \in \bar{N}'_s, t \in \bar{T} \tag{32}$$

Constraints (31) guarantee that if a part-time interpreter is not assigned to any LEP patient, then she is not hired. Constraints (32) ensure that a part-time interpreter is hired even if she is assigned only once over all scenarios $S$ to an LEP patient.

### 4.1.4 Simplified and strengthened version of the SAA model

$$\text{Min} \sum_{j=1}^{J'} F_j^p w_j + \frac{1}{S} \sum_{s=1}^{S} \left( \sum_{j=1}^{J'} C_j^p ST_{js} + \sum_{i=1}^{I'} OC_i^f OT_{is} + \sum_{n=1}^{N'_s} P_n U_{ns} \right) \tag{33}$$

Subject to:

$$\sum_{i=1}^{I'} \sum_{t=1}^{T} x_{ints} + \sum_{j=1}^{J'} \sum_{t=1}^{T} y_{jnts} \leq 1 \quad \forall n \in \bar{N}'_s, s \in S \tag{34}$$

$$\sum_{n=1}^{N'_s} x_{ints} \leq A_{it}^f \quad \forall i \in \bar{I}', t \in \bar{T}, s \in S \tag{35}$$

$$\sum_{n=1}^{N'_s} y_{jnts} \leq A_{jt}^p \quad \forall j \in \bar{J}', t \in \bar{T}, s \in S \tag{36}$$

$$x_{in'hs} \leq (1 - x_{ints}) \quad \forall i \in \bar{I}', t \in \bar{T}, s \in S, n \in \bar{N}2, n' \in \bar{N}'_s \setminus \{n\}, h \in \{t+1, \ldots, \min(T, t + D2_{ns} - 1)\} \tag{37}$$

$$x_{in'hs} \leq (1 - x_{ints}) \,\forall\, i \in \bar{I}', t \in \bar{T}, s \in S, n \in \bar{N}3_s, n' \in \bar{N}'_s\backslash\{n\}, h \in \{t+1,\ldots,\min(T, t+D3_{ns}-1)\} \tag{38}$$

$$y_{jn'hs} \leq (1 - y_{jnts}) \,\forall\, i \in \bar{J}', t \in \bar{T}, s \in S, n \in \bar{N}2, n' \in \bar{N}'_s\backslash\{n\}, h \in \{t+1,\ldots,\min(T, t+D2_{ns}-1)\} \tag{39}$$

$$y_{jn'hs} \leq (1 - y_{jnts}) \,\forall\, i \in \bar{J}', t \in \bar{T}, s \in S, n \in \bar{N}3_s, n' \in \bar{N}'_s\backslash\{n\}, h \in \{t+1,\ldots,\min(T, t+D3_{ns}-1)\} \tag{40}$$

$$\sum_{i=1}^{I'} \sum_{t=1}^{T} t x_{ints} + \sum_{j=1}^{J'} \sum_{t=1}^{T} t y_{jnts} + \alpha T(1 - \sum_{i=1}^{I'} \sum_{t=1}^{T} x_{ints} - \sum_{j=1}^{J'} \sum_{t=1}^{T} y_{jnts}) - t2_n \leq U_{ns} \,\forall s \in S, n \in \bar{N}2 \tag{41}$$

$$\sum_{i=1}^{I'} \sum_{t=1}^{T} t x_{ints} + \sum_{j=1}^{J'} \sum_{t=1}^{T} t y_{jnts} + \alpha T(1 - \sum_{i=1}^{I'} \sum_{t=1}^{T} x_{ints} - \sum_{j=1}^{J'} \sum_{t=1}^{T} y_{jnts}) - t3_{sn} \leq U_{ns} \,\forall s \in S, n \in \bar{N}3_s \tag{42}$$

$$x_{ints} = 0 \,\,\forall\, i \in \bar{I}', n \in \bar{N}2, t < t2_n, s \in S \tag{43}$$

$$y_{jnts} = 0 \,\,\forall\, j \in \bar{J}', n \in \bar{N}2, t < t2_n, s \in S \tag{44}$$

$$x_{ints} = 0 \,\,\forall\, i \in \bar{I}', n \in \bar{N}3_s, t < t3_{sn}, s \in S \tag{45}$$

$$y_{ints} = 0 \,\,\forall\, j \in \bar{J}', n \in \bar{N}3_s, t < t3_{sn}, s \in S \tag{46}$$

$$x_{ints} \leq z_{in}^{f} \,\,\forall\, i \in \bar{I}', n \in \bar{N}2, t \in \bar{T}, s \in S \tag{47}$$

$$y_{jnts} \leq z_{jn}^{p} \,\,\forall\, j \in \bar{J}', n \in \bar{N}2, t \in \bar{T}, s \in S \tag{48}$$

$$x_{ints} \leq z3_{ins}^{f} \,\,\forall\, i \in \bar{I}', n \in \bar{N}3_s, t \in \bar{T}, s \in S \tag{49}$$

$$y_{jnts} \leq z_{jns}^{p} \,\,\forall\, j \in \bar{J}', n \in \bar{N}3_s, t \in \bar{T}, s \in S \tag{50}$$

$$\sum_{n=1}^{N2} \sum_{t=1}^{T} x_{ints} D2_{sn} + \sum_{n=N2+1}^{N's} \sum_{t=1}^{T} x_{ints} D3_{sn} = H_{is}^{f} \,\,\forall\, i \in \bar{I}', s \in S \tag{51}$$

$$\sum_{n=1}^{N2} \sum_{t=1}^{T} y_{jnts} D2_{sn} + \sum_{n=N2+1}^{N's} \sum_{t=1}^{T} y_{jnts} D3_{sn} = H_{js}^{p} \,\,\forall\, j \in \bar{J}', s \in S \tag{52}$$

$$w_j \geq \varepsilon \sum_{s=1}^{S} \sum_{n=1}^{N'_s} \sum_{t=1}^{T} y_{jnts} \,\,\forall j \in \bar{J}' \tag{53}$$

$$H_{is}^{f} - WR_i^{f} \leq OT_{is} \,\forall\, i \in \bar{I}', s \in S \tag{54}$$

$$H_{js}^{p} - WF_j^{p} \leq ST_{js} \,\forall j \in \bar{J}', s \in S \tag{55}$$

$$w_j \leq \sum_{s=1}^{S} \sum_{n=1}^{N'_s} \sum_{t=1}^{T} y_{jnts} \,\,\forall j \in \bar{J}' \tag{56}$$

$$w_j \geq y_{jnts} \,\,\forall\, j \in \bar{J}', s \in \{1..S\}, n \in \bar{N}'_s, t \in \bar{T} \tag{57}$$

$$w_j, x_{ints}, y_{jnts} \in \{0,1\} \,\,\forall\, j \in \bar{J}' \,\forall\, i \in \bar{I}', n \in \bar{N}'_s, t \in \bar{T}, s \in S \tag{58}$$

$$U_{ns}, H^f_{is}, H^p_{js}, OT_{is}, ST_{js} \geq 0, \forall n \in \bar{N}'_s, \forall i \in \bar{I}', \forall j \in \bar{J}', s \in S \tag{59}$$

Constraints (34)-(55) are the deterministic equivalent of constraints (2)-(4), (6)-(7), (9)-(10), (12)-(13), (16)-(28), respectively, after model simplification and the subsequent adjustment of sets and their cardinals described above. Constraints (56) and (57) are the introduced valid cuts while constraints (58) and (59) are domain constraints.

### 4.1.5 SAA Algorithm

The procedure of the used SAA algorithm is as follows:

1. For $m=1..M$ do steps 1.1 through 1.4

    Step 1.1 Generate a sample of size $S$

    Step 1.2 Solve the corresponding SAA model (i.e., model (33)-(59)) and save its optimal solution $\widehat{w}^m_S$ and the corresponding optimal objective value $\hat{o}^m_S$

    Step 1.3 Generate a sample of size $S'$ ($S'>S$)

    Step 1.4 Using the generated sample of size $S'$, estimate the "true" objective value $\hat{g}_{S'}(\widehat{w}^m_S)$ of the SAA optimal solution $\widehat{w}^m_S$ and the corresponding variance $\sigma^2_{\hat{g}_{S'}(\widehat{w}^m_S)}$ by (60) and (61)

$$\hat{g}_{S'}(\widehat{w}^m_S) = \frac{1}{S'}\sum_{s=1}^{S'}\hat{g}_s(\widehat{w}^m_S) \tag{60}$$

$$\sigma^2_{\hat{g}_{S'}(\widehat{w}^m_S)} = \frac{1}{(S'-1)S'}\sum_{s=1}^{S'}(\hat{g}_s(\widehat{w}^m_S) - \hat{g}_{S'}(\widehat{w}^m_S))^2 \tag{61}$$

2. Estimate $\bar{o}^M_S$ and the corresponding variance $\sigma^2_{\bar{o}^M_S}$ by (62) and (63)

$$\bar{o}^M_S = \frac{1}{M}\sum_{m=1}^{M}\hat{o}^m_S \tag{62}$$

$$\sigma^2_{\bar{o}^M_S} = \frac{1}{(M-1)M}\sum_{m=1}^{M}(\hat{o}^m_S - \bar{o}^M_S)^2 \tag{63}$$

3. Select the best solution $\widehat{w}^{m*}_S$, $m* = \underset{m=1..M}{\operatorname{argmin}}\widehat{w}^m_S$, and compute the corresponding optimality gap $(\hat{g}_{S'}(\widehat{w}^{m*}_S) - \bar{o}^M_S)$ and the corresponding variance $\sigma^2_{gap}$ by (64)

$$\sigma^2_{gap} = \sigma^2_{\bar{o}^M_S} + \sigma^2_{\hat{g}_{S'}(\widehat{w}^{m*}_S)} \tag{64}$$

At this level, it is worth noting that $\bar{o}^M_S$ and $\hat{g}_{S'}(\widehat{w}^{m*}_S)$ are a statistical lower and upper bound of the optimal objective function of the original two-stage stochastic programming model (Mak et al. 1999). A $(1 - 2\alpha)$-confidence interval for the optimality gap of the solution $\widehat{w}^{m*}_S$ is given by (65)

$$(\hat{g}_{S'}(\widehat{w}^{m*}_S) - \bar{o}^M_S) \pm Z_\alpha \sigma_{gap} \tag{65}$$

With $Z_\alpha = \Phi^{-1}(1 - \alpha)$, where $\Phi$ is the cumulative distribution function of the standard normal distribution.

For each solution obtained through the above procedure, we should examine the value of its optimality gap and corresponding variance. If these values are too large, one must repeat the procedure with larger values of *S* and/or *M*. For more details on the SAA algorithm, we refer the readers to (Kleywegt et al. 2001).

## 4.2 Tabu Search (TS) algorithm

We now introduce the developed TS algorithm to solve the SAA model in a more reasonable computational time. At this level, it is worth noting that TS is a neighborhood-based metaheuristic that proves effective in solving deterministic personnel scheduling problems (Van den Bergh et al. 2013). The developed TS embeds Monte Carlo sampling, and a heuristic for the evaluation of the second-stage objective function, which will be hereafter referred to as HESTOF. Obviously, HESTOF includes the construction of the interpreter schedule for each considered scenario.

Figure 1 illustrates the steps of the TS algorithm. The array bearing the first-stage solution of the SAA model, i.e., the hiring decision $w$ ($w_1, \ldots w_{J'}$), is encoded in a way that part-time interpreters are sorted according to the language or the set of languages, for which they can provide interpreting service. As can be seen from Figure 2, the interpreters who can provide interpreting services for a single language, such as Spanish or Russian, are placed first, while those who can interpret for many languages are placed last. The set of interpreters who can provide service for the same language(s) can be, therefore, easily distinguished from the array. First, an initial solution ($w_{init}$) is generated and set as the current solution ($w_{cur}$). The initial solution ($w_{init}$) is produced by solving the SAA model while incorporating a small number of scenarios (for instance, one, five, etc.). Obviously, the number of scenarios for solving the SAA model is chosen in such a way that the associated computational burden remains below a certain threshold. Then, at each iteration, *N* solutions in the neighborhood of the current solution ($w_{cur}$) are generated. The fitness of each generated solution, i.e., its objective function value, is evaluated using HESTOF. The best solution among these *N* candidate solutions could be selected to become the new current solution even if it is worse than the current best solution. To prevent cycling, TS keeps track of visited solutions by storing them in a queue, called tabu list (TL) (Tsai and Chiang 2023). A candidate solution can be explored by the search only if it is not in the TL. Moreover, a visited solution remains in the TL for a certain time that depends on its length. Indeed, whenever the TL is full, the oldest solution is removed. As such, in each iteration, the TL allows to filter out the solutions in the neighborhood of the current solution that have been recently visited. The best candidate solution over those not in the TL is selected; it becomes the new current solution ($w_{cur}$) and is added to the TL. This is repeated until a stopping criterion is reached. At that point, the best-found solution along the search is returned. In the following, we describe the used neighborhood structure and diversification strategy. Also, we present HESTOF.

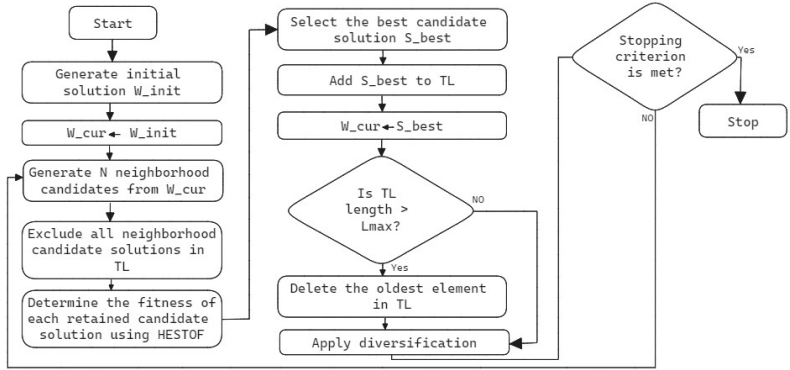

Figure 1. Flow chart of the proposed TS algorithm

| Interpreter index | 1 | 2 | 3 | 4 | 5 | .... | J'-5 | J'-4 | J'-3 | J'-2 | J'-1 | J' |
|---|---|---|---|---|---|---|---|---|---|---|---|---|
| w | 1 | 0 | 1 | 0 | 1 | .... | 1 | 0 | 0 | 0 | 1 | 0 |
| | | Language 1 | | Language 2 | | | Language 1 and 2 | | | Language 1, 2 and 3 | | |

Figure 2. Representation of the first-stage solution in the proposed TS

4.2.1 Neighborhood structure

The TS employs a two-swap heuristic to explore the neighboring solutions of a current solution ($w_{cur}$). Two-swap is first applied separately on each of the set of successive elements in the solution array corresponding to the set of interpreters who can provide the same service (i.e., interpret for the same languages). Let $S_1, S_2, \ldots S_L$ represent these sets. For each set $S_l$ ($l=1..L$), the two-swap heuristic randomly selects two elements $i$ and $j$ and swaps their respective values. The new values of $S_1, S_2, \ldots S_L$ are then concatenated together to form a new candidate solution (see Figure 3). Following this scheme, $N$ neighborhood solutions are generated in each iteration. To ensure that the final set of candidate solutions is non-redundant, the $N$ candidate solutions are checked, and any duplicates are removed.

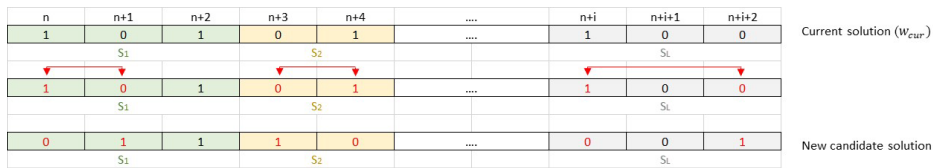

Figure 3. Two-swap heuristic

4.2.2 Diversification strategy

Diversification is needed by the TS to favor the exploration of new regions of the search space. In the proposed TS, a diversification strategy is applied, with a pre-defined probability, by generating a completely new solution using again SAA with a small number of scenarios based on a small probability (see figure 1). The generated solution will be therefore considered as the new current solution ($w_{cur}$).

### 4.2.3 HESTOF

As indicated above, HESTOF is a heuristic that builds interpreter schedule for a given candidate solution (i.e., a hiring decision $w$ ($w_1,...w_{J'}$)). As such, embedded in the TS, it allows to evaluate the fitness (i.e., the objective function value) of a candidate solution based on a set of generated scenarios.

Let us introduce some notation used in HESTOF. $A$ is the set containing the times over which each interpreter becomes available. $A$ includes both full and part-time interpreters and is ordered following the same rule used for ordering the elements of the array $w$. $A$ will be updated whenever an interpreter is assigned to an LEP patient. $R$ is the set of arrival times of LEP patients (outpatients and emergency patients) ordered in increasing order. $NS$ is the set of patients not served. Initially, $NS$ contains all the LEP patients (outpatients and emergency patients). It is updated whenever an interpreter is assigned to an LEP patient. As can be noticed, the sets $R$ and $NS$ are scenario dependent. For that, HESTOF starts by generating a sample of $S$ scenarios. For each scenario, the sets $R$ and $NS$ are determined, as well as the service time for each LEP patient. For simplification, we did not incorporate a scenario index and will present HSTOF when applied to one scenario, referred to as HSTOF_1 (see Algorithm 1). The presented steps will be applied to each scenario among the $S$ scenarios. After that, the expected objective function value of a candidate solution $w$ was evaluated based on the $S$ scenarios.

---
**Algorithm 1: HSTOF_1**

Initial input: $A, R, NS$
For $t$=1 to $T$ do
**Begin**
  If ($t \in R \cup A$) then
  **Begin**
    Determine $P_t$ the set of LEP patients waiting for service at time $t$
    Select the set of languages needed by the LEP patients in $P_t$
    Select the ordered set $I_t$ of available interpreters that can ensure the service for those languages
    //according to their order in the array $w$
    For each interpreter $i$ in $I_t$ do
    **Begin**
      Select $Y_t$, the subset of LEP patients in $P_t$, for which she can ensure interpreting service
      For each patient $n$ in $Y_t$ do
      **Begin**
        Evaluate the average penalty cost for waiting associated with choosing patient $n$
        Select the patient $n'$ with the lowest average penalty cost and assign it to interpreter $i$
        $t_{n'}=t$
        // $t_{n'}$ is the start time for service for patient $n'$
        Update the sets $A$, $NS$ and $P_t$
      **End**
    **End**
  **End**
**End**
Determine the objective function value of the constructed schedule

## 5. Experimentation and numerical results

First, we introduce the real-life case study that has been used in the numerical experimentation. Then, we present and compare the results obtained from the SAA and the TS algorithms. Next, we conduct a sensitivity analysis by investigating the effect of some parameters on part-time interpreter hiring decisions, the costs and the QoS. Thereafter, these results are used to draw valuable managerial insights.

The proposed SAA and TS algorithms have been implemented using Python. All the MIPs of the SAA algorithms are solved using Gurobi Optimizer (version 9.5.0), with all parameters set to their default values. Moreover, all experiments were conducted on the University of Alabama at Birmingham's High-Performance Computing (HPC) server, deploying 128 physical cores and 21 GB of RAM.

### 5.1 Data description

Our experimentation is based on real-life data that were collected from our partner hospital, which is a medium-sized hospital with about 350 beds and 300 nurses and physicians. The hospital provides medical services for about 55,000 patients per year, including a significant proportion of patients with LEP. The part-time interpreters are hired according to the need of the hospital, based on a contract between the hospital and an interpreting agency. By the established contract, at the end of each day, the hospital should confirm the interpreters to be hired for the following day. The contract entails a payment structure comprising fixed and variable costs. The fixed cost is charged regardless of how long the interpreting service time is. It covers the cost of service for a certain agreed-upon time threshold. Beyond this time threshold, the pay of a part-time interpreter involves a variable cost.

Although the proposed approach can handle any number of languages, we consider the five most interpreted languages, which are Hmong, Spanish, Somalian, Vietnamese, and Russian. At the partner hospital, an interpreter is assigned to each LEP patient upon admission. The latter provides the interpreting service to the LEP inpatient each day of their stay (specifically, between 8 am and 5 pm). Outside these hours of the day, the interpreting service is offered by phone. In general, full-time interpreters are assigned to LEP inpatients. Part-time interpreters are only assigned to LEP inpatients if the number of available full-time interpreters is not enough to cover the service. For the day considered in the experimentation, there are six LEP inpatients (four needing Spanish interpreting service, one needing Vietnamese interpreting service and one needing Hmong interpreting service) and 14 LEP outpatients. As indicated in subsection 4.1.2, the six LEP inpatients and the full-time interpreters assigned to them are not included in the SAA model. The planning horizon is 8 hours, encompassing 24 periods of 20 minutes each.

The stochastic parameters include the number of arriving emergency patients with LEP and the duration of interpreting sessions for both outpatients and emergency patients. Table 1 reports their probability distributions and associated parameters, which are fitted from the data collected from the partner hospital. For each interpreting service (i.e., language), the arrivals of emergency patients with LEP were fitted into a Poisson distribution. For the service time of outpatients, different distributions were tested, then the best-fit distribution was selected, which was the normal distribution. For the emergency patients with LEP, since getting data for the service time was not feasible, uniform distribution was assumed based on the opinion of an expert who is a full-time interpreter at the partner hospital. Moreover, Table 1 presents the number of available full-time and part-time interpreters, along with the values of the cost parameters extracted from the collected data.

Table 1. Values of the used parameters

| Parameter category | Parameter | Value |
|---|---|---|
| Full-time Interpreters (12) | # for Hmong language | 6 |
| | # for Russian language | 1 |
| | # for Somalian language | 1 |
| | # for Vietnamese language | 1 |
| | # for Spanish language | 3 |
| | Overtime rate | $13-$17 per period |
| Part-time interpreters (22) | # Hmong language | 6 |
| | # Russian language | 2 |
| | # Somalian language | 1 |
| | # Vietnamese language | 2 |
| | # Spanish language | 10 |
| | Fixed cost | $40 - $ 60 |
| | Covered time threshold | 8 periods |
| | Variable cost | $7 - $ 10 per period |
| LEP outpatients | Arrival | Deterministic |
| | Service time | Normal distribution (7.20, 5.25) |
| | Penalty waiting cost | $15 per period |
| Emergency patients with LEP | Arrival | Poisson distribution<br>Hmong ($\lambda = 0.13$) per period<br>Russian ($\lambda = 0.013$) per period<br>Somalian ($\lambda = 0.050$) per period<br>Vietnamese ($\lambda = 0.022$) per period<br>Spanish ($\lambda = 0.184$) per period |
| | Service time | Uniform Distribution (6, 12) |
| | Penalty waiting cost | $30 per period |

## 5.2 Numerical results

This section evaluates the performance of the SAA and the TS algorithms using the base case. Therefore, for the two algorithms, the quality of the obtained solutions and their computational times are assessed.

### 5.2.1 Results of the SAA Algorithm

The SAA algorithm is implemented with $M = 5$, $S' = 500$, and progressively increasing the sample size $S$ from 5 to 100 (namely, with $S = 5, 10, 20, 30, 50$, and $100$). Table 2 and 3 report the obtained results. The first column in Table 2 provides the sample size considered while solving the SAA problem. Column two reports the average computational time needed to solve the SAA model. Columns three through six delineate the statistical lower and upper bounds (referred to as LB and UB, respectively) and the estimates of their standard deviations.

Table 2. Statistical lower and upper bounds of the SAA problems for *M*=5 and *S'*=500

| Sample Size S | Avg. CPU (min) | Lower Bound | | Upper Bound | |
|---|---|---|---|---|---|
| | | LB | Std. | UB | Std. |
| 5 | 5.20 | 657.17 | 32.26 | 915.54 | 25.06 |
| 10 | 13.33 | 684.92 | 19.88 | 890.84 | 26.69 |
| 20 | 42.34 | 723.99 | 12.51 | 767.91 | 14.51 |
| 30 | 98.23 | 749.74 | 9.50 | 748.68 | 9.24 |
| 50 | 289.29 | 754.04 | 9.89 | 747.97 | 8.76 |
| 100 | 2033.21 | 758.11 | 5.67 | 748.53 | 8.99 |

Table 2 shows that, in general, when the sample size increases, the lower bound increases while the upper bound decreases. Furthermore, the associated standard deviations generally decrease when the sample size increases. Nevertheless, the time required to solve the SAA problem increases with the sample size, reaching a maximum average of approximately 33 hours for a sample size of 100.

Table 3 provides more details on the quality of the best solutions found. Columns two through four present the estimates of the optimality gap (in value and percentage) of the best solution found, as well as the standard deviation of the gap. Columns five through eight report the 95% confidence intervals of the optimality gap (in value and percentage). Figure 4 portrays the width of the 95% confidence interval of the optimality gap.

Table 3 and Figure 4 demonstrate the convergence of the SAA algorithm. Indeed, as can be seen, when the sample size increases, the 95% confidence interval of the optimality gap associated with the best-obtained solution becomes narrower. Therefore, considering samples with a larger number of scenarios in the SAA algorithm can provide a stronger guarantee regarding the closeness of the found solutions to the optimal one. Markedly, the optimality gap becomes very small, starting from a sample comprising 30 scenarios. It can also be noted that for some sample sizes, the estimate of the optimality gap is negative. This is not unexpected given that the lower and the upper bounds obtained by the SAA algorithm, and so the optimality gap, are statistical in nature. However, since these optimality gaps are not smaller than minus twice their respective standard deviations, they are accepted (Verweij et al. 2003).

From Figure 4, one can also see that a sample of size $S = 30$ offers a good trade-off between the quality of the solution and the computational effort needed to obtain it. Indeed, with $S = 30$, a near-optimal solution is obtained within a much more reasonable computational time (with an average of 98.23 minutes) compared to the time required to obtain a solution with a comparable optimality gap using a sample size of $S = 100$ (with an average of 2033.21 minutes). In particular, this result highlights the applicability of the proposed SAA scheme to the considered interpreter scheduling problem.

Table 3. Optimality gap and corresponding 95% confidence interval with $M=5$ and $S'=500$

| Sample Size $S$ | Optimality gap | | | 95% confidence interval | | | |
|---|---|---|---|---|---|---|---|
| | Estimate | % | Std. | Min. | % | Max. | % |
| 5 | 258.37 | 39.31 | 40.85 | 178.30 | 27.13 | 338.44 | 51.50 |
| 10 | 205.92 | 30.06 | 33.28 | 140.69 | 20.54 | 271.16 | 39.59 |
| 20 | 43.92 | 6.07 | 19.16 | 6.37 | 0.88 | 81.47 | 11.25 |
| 30 | -1.06 | -0.14 | 13.25 | -27.03 | -3.60 | 24.91 | 3.32 |
| 50 | -6.07 | -0.80 | 13.21 | -31.96 | -4.23 | 19.82 | 2.63 |
| 100 | -9.58 | -1.26 | 10.63 | -30.41 | -4.01 | 11.25 | 1.48 |

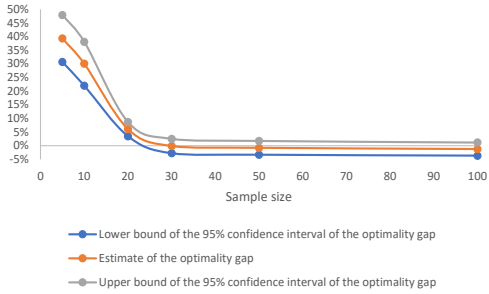 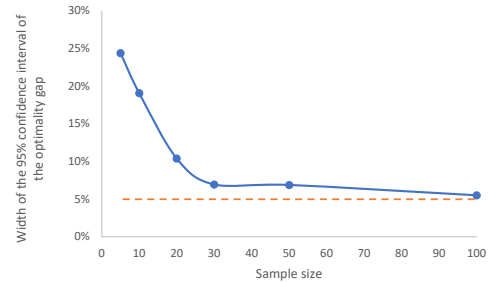

(a) 95% confidence interval for the optimality gap     (b) Width of the 95% confidence interval

Figure 4. 95% confidence interval for the optimality gap and its width

### 5.2.2 Results of the TS Algorithm

First, preliminary tests are conducted to select the parameters of the TS algorithm, namely the number of iterations, the probability of diversification, and the length of the tabu list. For instance, the TS algorithm considering one scenario was run while setting the number of iterations to 300, 500, and 1000. It was found that the algorithm converged before reaching 100 iterations. Therefore, 100 was the number of iterations selected for the TS algorithm. Similarly, several values for the length of the tabu list and the probability of diversification were tested, and those indicated in Table 4 were retained for the algorithm.

Table 4. Parameters of the TS algorithm

| Parameter | Value |
|---|---|
| Number of iterations | 100 |
| Probability of diversification | 0.05 |
| Length of the tabu list | 30 |

The TS algorithm is evaluated while considering two sample sizes, namely 50 and 100 scenarios. For each sample size, the algorithm is run while incorporating initial solutions obtained by solving the SAA model with a sample including one, five, or ten scenarios. For each setting, the TS algorithm is run 5 times. Each obtained solution is then simulated by considering a sample of 500 scenarios. The solution with the lowest objective function value, as revealed by the simulation, is selected. For each setting (i.e., sample size in the initial solution (Sample size_IS) and sample size considered in the TS algorithm (Sample size_TS)), Table 5 reports the average computational time, the estimate of the objective function value of the selected solution (by the TS algorithm (Best_TS), and by simulation (UB)) and the estimate of its standard deviation. Table 5 and Figure 5 demonstrate that the best solution is obtained when a sample of size 100 is used in the TS algorithm. Moreover, as expected, the average computational time is higher when a sample of size 100 is used. One can also note from Table 4 that the sample size used to obtain the initial solution does not clearly affect the quality of the obtained solution.

Table 5. Results of the TS algorithm

|   | Sample size_IS | Sample size_TS | Best_TS | Avg. CPU (min) | Upper Bound | |
|---|---|---|---|---|---|---|
|   |   |   |   |   | UB | Std. |
| 1 | 1 | 50 | 784.31 | 20.23 | 849.50 | 14.52 |
| 2 | 1 | 100 | 811.13 | 25.65 | 818.49 | 12.15 |
| 3 | 5 | 50 | 800.68 | 17.14 | 961.06 | 10.64 |
| 4 | 5 | 100 | 804.10 | 23.55 | 784.48 | 8.04 |
| 5 | 10 | 50 | 814.38 | 16.30 | 939.59 | 8.13 |
| 6 | 10 | 100 | 828.04 | 23.2 | 787.18 | 8.07 |

The optimality gap of the selected solutions and the corresponding 95% confidence intervals are reported in Table 6. It is noteworthy that, for each selected solution, the optimality gap is estimated using the best lower bound from the SAA algorithm. The standard deviation of the optimality gap and the 95% confidence interval are determined using (64) and (65), respectively, using the standard deviation of the best lower bound.

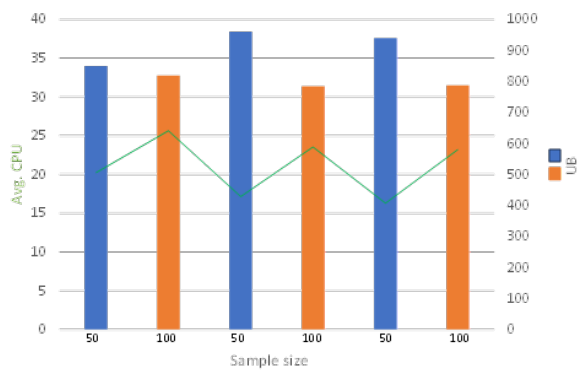

Figure 5. Results of the TS algorithm

Table 6. Optimality gap and corresponding 95% confidence interval of the TS algorithm solutions

|   | Optimality gap | | | 95% confidence interval | | | |
|---|---|---|---|---|---|---|---|
|   | Estimate | % | Std. | Min. | % | Max. | % |
| 1 | 91.39 | 12.06 | 15.59 | 60.83 | 8.024 | 121.95 | 16.08 |
| 2 | 60.38 | 7.96 | 13.41 | 34.09 | 4.49 | 86.66 | 11.43 |
| 3 | 202.95 | 26.77 | 12.05 | 179.33 | 23.65 | 226.57 | 29.88 |
| 4 | 26.37 | 3.48 | 9.84 | 7.08 | 0.93 | 45.66 | 6.02 |
| 5 | 181.48 | 23.94 | 9.91 | 162.06 | 21.38 | 200.90 | 26.50 |
| 6 | 29.07 | 3.83 | 9.86 | 9.74 | 1.28 | 48.39 | 6.38 |

The widths of the 95% confidence intervals of the optimality gaps of the solutions obtained with a sample of size 100 are generally smaller than those obtained with a sample of size 50. As such, considering samples with a larger number of scenarios in the TS algorithm allows us to obtain solutions of higher quality. The optimality gap of the best TS algorithm solution is 3.48%. More importantly, this near-optimal solution was obtained within 26 minutes. Therefore, compared to the SAA algorithm with a sample size of 30, the TS algorithm can provide near-optimal solutions within a much more reasonable computational time. Although the quality of the solution might be slightly compromised when the TS algorithm is used (according to the 95% confidence interval of the optimality gap), the solution remains near-optimal, and the computational burden is significantly reduced.

5.2.3 Comparison of different solutions

In order to assess the importance of addressing the stochasticity of the demand and the service time, we compare the solutions obtained from the SAA and the TS algorithms with the solution of the Expected Value Problem (EVP), in which stochastic parameters are replaced by their expected values. Figure 6 portrays the average and the standard deviation of the total cost assessed for the three solutions mentioned above using simulation. It clearly demonstrates the successive improvement in total cost when adopting a deterministic approach, a stochastic programming approach using the TS algorithm, and an approach using the SAA algorithm. The total cost is reduced by 19.87% when SAA algorithm is used and by 16.04% when TS algorithm is used. It can also be observed that the number of hired part-time interpreters in the EVP solution is lower than the one proposed by the SAA and the TS algorithm solutions. Hiring additional part-time interpreters incurs higher fixed costs but significantly reduces the penalty cost for waiting. The average total waiting time is reduced from 6 periods to around 1 period when a stochastic programming approach is used. Furthermore, the service level for emergency patients with LEP increased from 91.88% to above 99.90%. As such, very interestingly, the use of a stochastic programming approach for the scheduling of interpreting services improves both the total cost and the QoS.

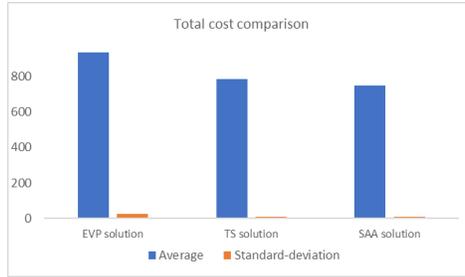

Figure 6. Cost comparison of different solutions

## 5.3 Sensitivity analysis

Sensitivity analysis is carried out to investigate the effect of five problem parameters, namely the unit overtime cost for full-time interpreters, the fixed cost and the unit pay rate (or variable cost) of part-time interpreters, the unit penalty cost for waiting, and the arrival rate of LEP emergency patients, on the hiring decision of part-time interpreters, the different involved costs, and the QoS. These parameters are increased one by one. In other words, the values of one parameter are increased while the other parameters are kept at their base values. The values of each parameter are increased incrementally by chunks of 20% until they are doubled, which results in conducting five experiments for each parameter, numbered from 2 to 6, respectively, in Figures 7 through 11. In these figures, portraying the effect of increasing the studied parameters, experiment 1 represents the results obtained for the base case.

In each experiment, first, the problem is solved using the TS algorithm to determine the number of part-time interpreters to be hired; then the true values of three performance measures are evaluated using simulation. For each experiment, as shown in Figures 7 through 11, in addition to the number of hired part-time interpreters, the following performance measures are reported: the distribution of the total cost, the average waiting time of LEP patients; and the average service level for emergency patients and outpatients. It can be noted that the last two performance measures pertain to QoS. The number of hired part-time interpreters is represented using a bar chart, in which the bars are colored to distinguish the interpreters hired per language. Similarly, the distribution of the total cost among the variable cost of part-time interpreters, the penalty cost for waiting, the overtime cost for full-time interpreters, and the fixed cost of part-time interpreters, is highlighted using different colors.

As can be observed from Figure 7, the total overtime cost is relatively small with comparison to the other costs. Therefore, an increase of the unit overtime cost has a little effect on the hiring decision, the cost distribution, the waiting time and the service level. Note that the latter remains above 99% in all experiments for LEP emergency patients and LEP outpatients. It can also be noted that the service level of emergency LEP patients is, in general, slightly higher than the one of the LEP outpatients. This can be explained by the higher priority given to LEP emergency patients, which is enforced by the associated higher unit penalty cost for waiting.

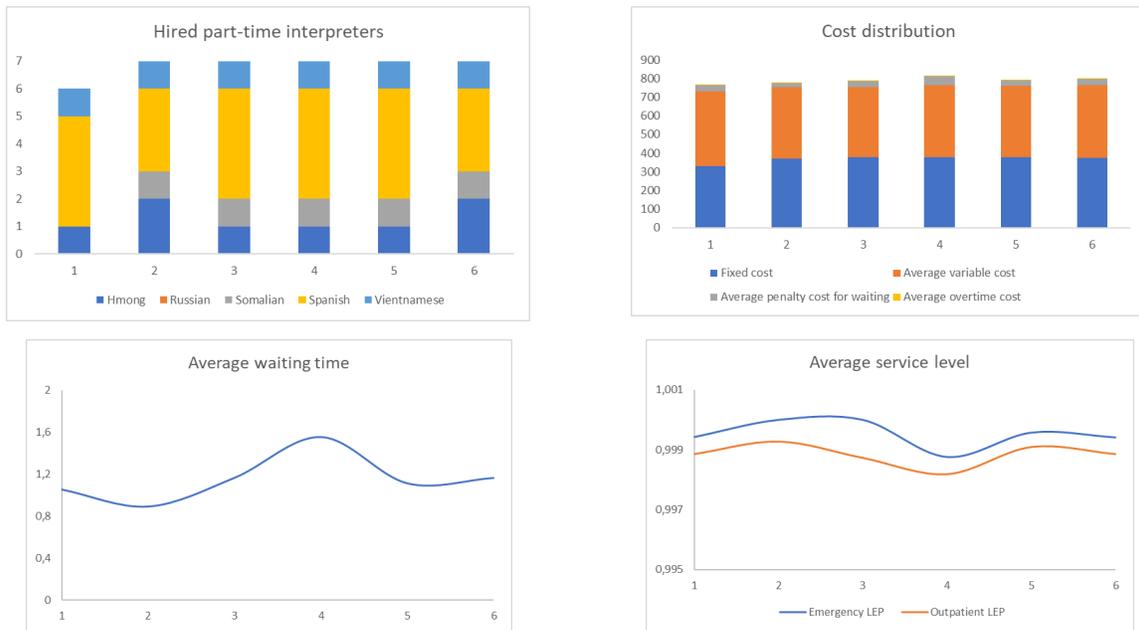

Figure 7. Effect of increasing unit overtime cost

Figure 8 shows that increasing the fixed cost associated with the hiring of part-time interpreters has no clear effect on the number of employed part-time interpreters as, in all experiments, 6 or 7 part-time interpreters are hired. However, it impacts the number of part-time interpreters hired for each language. Noticeably, the consistent hiring of part-time interpreters for Hmong and Spanish (across all experiments) can be explained by the relatively high number of outpatients and the high arrival rate of emergency patients with LEP requiring these languages. The exclusive hiring of part-time interpreters for Hmong and Spanish in experiment 6 is intended to limit the increase in the total fixed cost and was notably followed by a decrease in the variable cost. However, it results in an increase in the overtime cost for full-time interpreters and the penalty cost for waiting. Obviously, the increase in the penalty cost for waiting stems from an increase in the waiting time. The service level remains above 99%, meaning that while LEP patients are generally served, they tend to wait longer in this case.

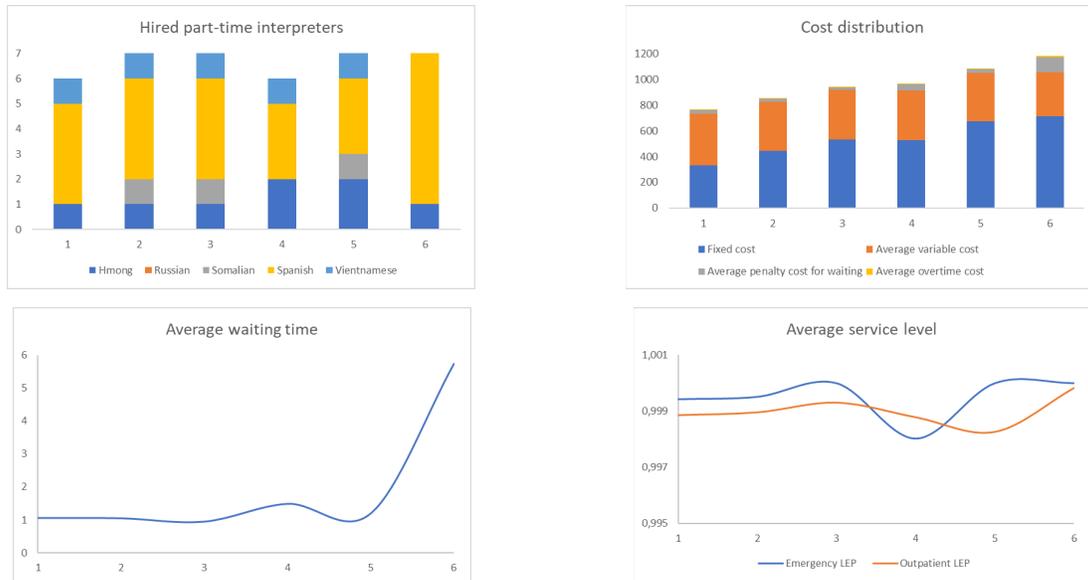

Figure 8: Effect of increasing the fixed cost of part-time interpreters

Figure 9 shows that the increase in the variable cost of part-time interpreters has no clear effect on the hiring decision. The number of hired part-time interpreters remains between 6 and 7. Therefore, as the unit pay rate of part-time interpreters increases, the total variable cost also increases. One can also note slight variations in waiting time and the service level, even in experiment 6, where the unit pay rate of part-time interpreters was doubled. Noticeably, in experiments 3-5, the service level for LEP outpatients is slightly higher than the one for LEP emergency patients, although both remain above 99%.

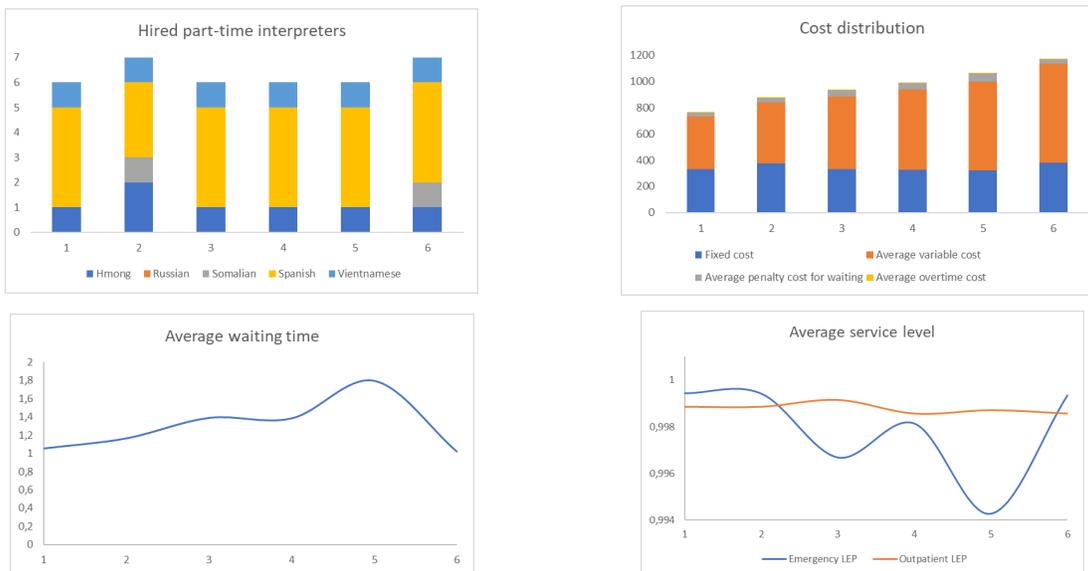

Figure 9: Effect of increasing the unit pay rate of part-time interpreters

Figure 10 shows the effect of increasing the unit penalty cost for waiting. As the penalty cost increases, the number of hired part-time interpreters would increase as this tends to reduce the waiting time. It is also worth noting that, in this case, the waiting time and the service level vary within narrow ranges as the unit penalty cost increases.

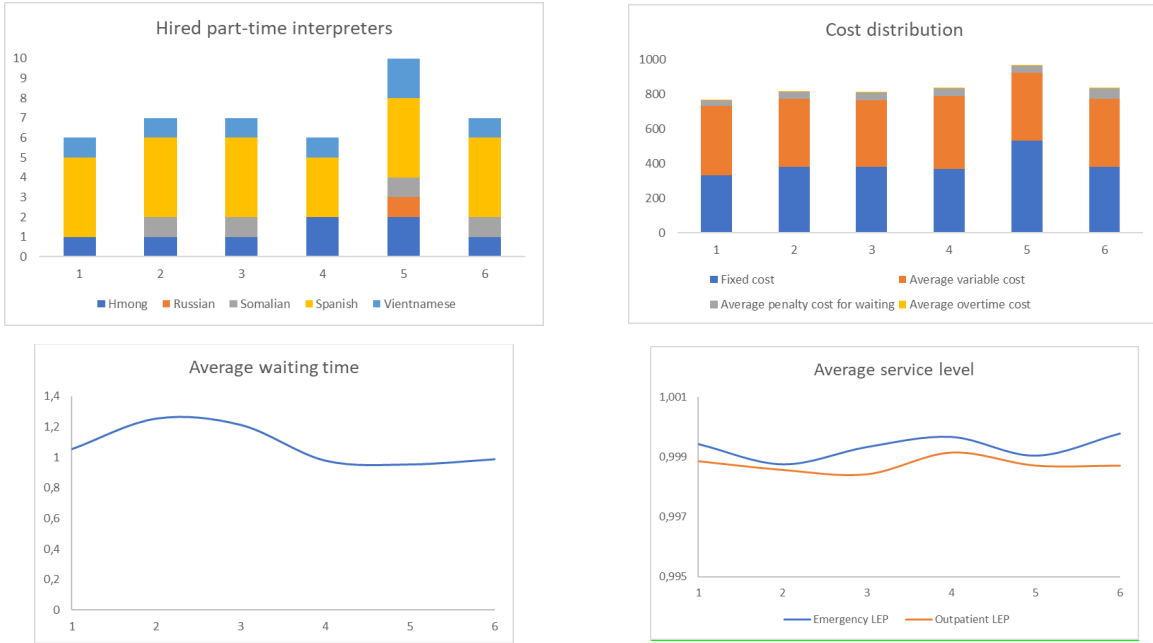

Figure 10: Effect of increasing the unit penalty cost for waiting

As can be noted from Figure 11, again here, when the arrival rate of emergency patients with LEP increased, the total number of hired part-time interpreters remains between 6 and 7. This decision however results in a notable increase in the total variable cost. In addition, the waiting time increased and so the associated penalty cost. The decision not to hire more part-time interpreters could be explained by the fact that the incurred cost is higher than the increase in the variable and the penalty cost for waiting. It can also be seen that when the demand from the LEP emergency patients increases, the corresponding service level dropped, reaching 94.1% in experiment 6. In addition, although priority is given to LEP emergency patients over outpatients, the service level for the former is lower than that achieved for the latter. This is due to the uncertainty in the arrival time of the LEP emergency patients. Indeed, the effect of this uncertainty intensifies as the number of emergency patients with LEP increases.

### 5.4 Managerial insights

The increase in the overtime cost, the variable cost of part-time interpreters and the arrival rate of emergency patients with LEP do not have a clear impact on the hiring decisions. The slight variation in the proposed solutions is incurred by the high uncertainty inherent to the considered problem. The increase in the fixed cost of hiring part-time interpreters could result in favoring the recruitment of part-time interpreters for the most demanded languages if the associated fixed costs are lower than the ones associated with the other interpreting services. An increase in the penalty cost for waiting favors the hiring of a larger number of part-time interpreters to avoid excessive waiting time and ensure higher service levels. Remarkably, the increase in the arrival rate of emergency patients with LEP can be followed by a decrease in their service level.

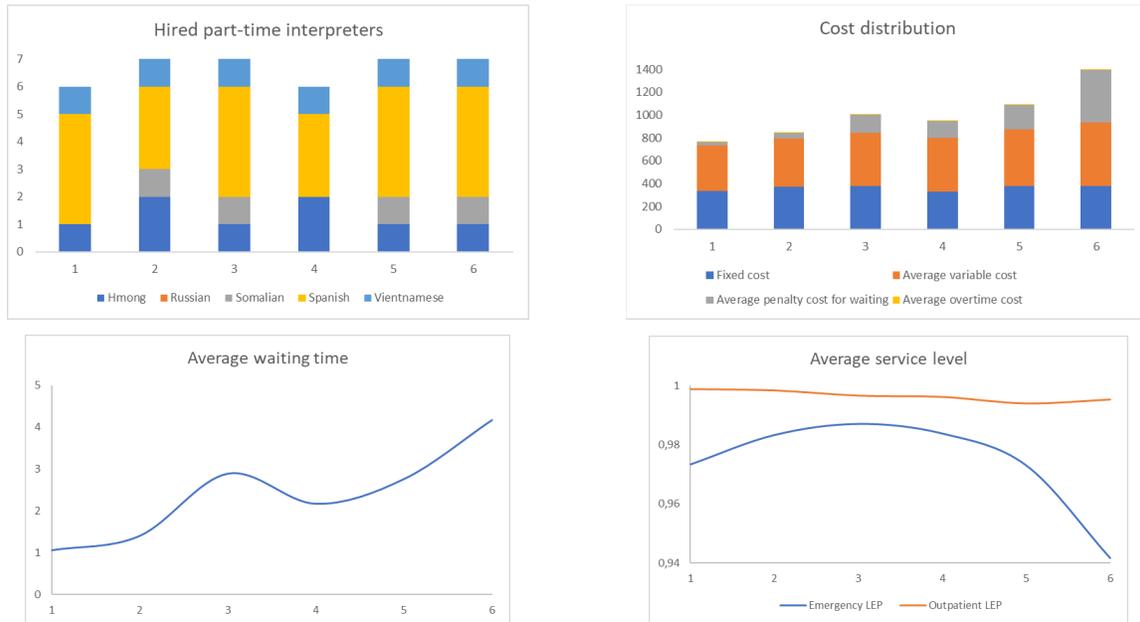

Figure 11: Effects of increasing the arrival rate of emergency patients with LEP

Overtime is seldom used in this case because its unit cost is higher than the unit variable cost of part-time interpreters. It is more advantageous to assign more LEP patients to part-time interpreters in scenarios involving a relatively high number of arriving emergency patients with LEP or the realization of longer session durations. As such, full-time interpreters are assigned to LEP patients in a manner that ensures their regular working hours are slightly exceeded. The rest of LEP patients are served by part-time interpreters.

The hiring of a larger number of part-time interpreters should ascertain a shorter waiting time and a higher service level. However, this was not opted for in the two cases where a large waiting time was observed (recall the two cases: (1) when the fixed cost of hiring part-time interpreters was doubled, and (2) when the arrival rate of emergency patients was doubled). This can be explained by the resulting high increase in the fixed cost because of the need to recruit several interpreters to ensure the interpreting service for many languages. The consideration in this hiring and scheduling problem of interpreters who can ensure service for several languages is therefore highly recommended. This allows more flexibility in delivering interpreting services, increase the efficiency and effectiveness of hiring decisions, and help reduce waiting times.

## 6. Conclusion and future work

In this paper, we investigated the daily scheduling of medical interpreting services that are provided to LEP patients. These services are ensured by full-time and part-time interpreters. For that, two decisions are considered: (1) the hiring of part-time interpreters and (2) the assignment of full-time and hired part-time interpreters to emergency patients and outpatients with LEP. First, the problem is formulated as a two-stage stochastic programming model to account for uncertainties related to the arrival of emergency patients and their service time, as well as the service time of outpatients. The minimization of the total cost, which incorporates full-time interpreters' overtime cost, the fixed and variable costs of part-time interpreters, and the penalty cost for not serving LEP patients on time, is sought. The model is solved using the SAA algorithm. Then to handle the computational burden of the SAA algorithm, the problem is solved using a well-designed TS algorithm. A real-life case study is used to validate the proposed solution algorithms and evaluate their performances. The results demonstrate the effectiveness of the SAA and the TS algorithm solutions in concurrently reducing the total cost and the waiting time compared to the Expected Value Problem (EVP) solution. Furthermore, the results ascertain the superior computational efficiency of the TS algorithm as it provides a near-optimal solution in less computational time.

A sensitivity analysis reveals that an increase in the fixed cost of hiring part-time interpreters could favor the recruitment of part-time interpreters for the most demanded languages if the associated fixed costs are relatively low. Moreover, an increase in the penalty cost for waiting generally favors the hiring of a larger number of part-time interpreters to avoid excessive waiting time and ensure higher service levels. An increase in the arrival rate of emergency patients with LEP triggers a slight drop in their service level.

In this work, we investigate the daily scheduling of medical interpreting service. Notably, this problem should be solved at the evening of each day to build the interpreters' schedule of the next day. Further work will address the online dynamic scheduling of interpreters happening over the course of a day. Information about the arrival of emergency patients with LEP, service requirements, service times and the availability of interpreters are dynamic and revealed in real-time. Appropriate models and advanced solution methods will be therefore developed to optimally solve the decision problem in a very short time frame.

## References

Abdalkareem, Z. A., A. Amir, M. A. Al-Betar, P. Ekhan, A. I. Hammouri. 2021. Healthcare scheduling in optimization context: a review. *Health Technol.* **11**(3): 445–469.

Afshar-Nadjafi, B. 2021. Multi-skilling in scheduling problems: A review on models, methods and applications. *Computers & Industrial Engineering* **151**: 107004.

Ahmed, A., Hamasha, M. M. 2018. Scheduling language interpreters at a medical center: An integer programming approach. *Proceeding of the 2018 Industrial and Systems Engineering Research Conference*, Orlando, FL, May 2018.

Ahmed, A., Frohn, E. 2021. A predictive and prescriptive analytical framework for scheduling language medical interpreters. *Health Care Manag Sci* **24**(3): 531–550.

Alvanchi, A., S. Lee, S. M. AbouRizk. 2012. Dynamics of workforce skill evolution in construction projects1This paper is one of a selection of papers in this Special Issue on Construction Engineering and Management. *Can. J. Civ. Eng.* **39**(9): 1005–1017.

Angelidis, E., D. Bohn, O. Rose. 2013. A simulation tool for complex assembly lines with multi-skilled resources. *2013 Winter Simulations Conference (WSC)*. 2577–2586.

Avramidis, A. N., W. Chan, M. Gendreau, P. L'Ecuyer, O. Pisacane. 2010. Optimizing daily agent scheduling in a multiskill call center. *European Journal of Operational Research* **200**(3): 822–832.

Bagheri, M., A. G. Devin, A. Izanloo. 2016. An application of stochastic programming method for nurse scheduling problem in real word hospital. *Computers & Industrial Engineering* **96**: 192–200.

Beagley, J., J. Hlavac, E. Zucchi. 2020. Patient length of stay, patient readmission rates and the provision of professional interpreting services in healthcare in Australia. *Health & Social Care in the Community* **28**(5): 1643–1650.

Bruni, R., P. Detti. 2014. A flexible discrete optimization approach to the physician scheduling problem. *Operations Research for Health Care* **3**(4): 191–199.

Çakırgil, S., E. Yücel, G. Kuyzu. 2020. An integrated solution approach for multi-objective, multi-skill workforce scheduling and routing problems. *Computers & Operations Research* **118**: 104908.

Castillo-Salazar, J. A., D. Landa-Silva, R. Qu. 2016. Workforce scheduling and routing problems: literature survey and computational study. *Ann Oper Res* **239**(1): 39–67.

Chiam, T. C., S. Hoover, D. Mosby, R. Caplan, S. Dolman, A. Gbadebo, F. Mayer, A. Nightingale, C.-A. Reyes-Hull, E. Brown. 2017. Meeting demand: A multi-method approach to optimizing hospital language interpreter staffing. *Journal of Hospital Administration* **6**(2): 21.

Cohen, A. L., F. Rivara, E. K. Marcuse, H. McPhillips, R. Davis. 2005. Are Language Barriers Associated With Serious Medical Events in Hospitalized Pediatric Patients? *Pediatrics* **116**(3): 575–579.

Divi, C., R. G. Koss, S. P. Schmaltz, J. M. Loeb. 2007. Language proficiency and adverse events in US hospitals: a pilot study. *International Journal for Quality in Health Care* **19**(2): 60–67.

Easton, F. F. 2011. Cross-training performance in flexible labor scheduling environments. *IIE Transactions* **43**(8): 589–603.

Easton, F. F. 2014. Service Completion Estimates for Cross-trained Workforce Schedules under Uncertain Attendance and Demand. *Production and Operations Management* **23**(4): 660–675.

Estrada, R. D., D. K. H. Messias. 2017. Language Co-Construction and Collaboration in Interpreter-Mediated Primary Care Encounters With Hispanic Adults. *Journal of Transcultural Nursing* 1043659617747523.

Flores, G., J. Rabke-Verani, W. Pine, A. Sabharwal. 2002. The importance of cultural and linguistic issues in the emergency care of children. *Pediatric Emergency Care* **18**(4): 271–284.

Fu, X., J. Qi, C. Yang, H. Ye. 2024. Elective Surgery Sequencing and Scheduling Under Uncertainty. *M&SOM* **26**(3): 893–910.

Gurazada, S. G., S. Gao, F. Burstein, P. Buntine. 2022. Predicting patient length of stay in Australian emergency departments using Data Mining. *Sensors* **22**(13): 4968.

Jacobs, B., A. M. Ryan, K. S. Henrichs, B. D. Weiss. 2018. Medical Interpreters in Outpatient Practice. *Annals of Family Medicine* **16**(1): 70.

Jacobs, E. A., D. S. Shepard, J. A. Suaya, E.-L. Stone. 2004. Overcoming Language Barriers in Health Care: Costs and Benefits of Interpreter Services. *Am J Public Health* **94**(5): 866–869.

John-Baptiste, A., G. Naglie, G. Tomlinson, S. M. H. Alibhai, E. Etchells, A. Cheung, M. Kapral, W. L. Gold, H. Abrams, M. Bacchus, M. Krahn. 2004. The Effect of English Language Proficiency on Length of Stay and In-hospital Mortality. *J Gen Intern Med* **19**(3): 221–228.

Ku, L., G. Flores. 2005. Pay now or pay later: providing interpreter services in health care. *Health Affairs* **24**(2): 435–444.

Kwan, M., Z. Jeemi, R. Norman, J. A. R. Dantas. 2023. Professional Interpreter Services and the Impact on Hospital Care Outcomes: An Integrative Review of Literature. *International Journal of Environmental Research and Public Health* **20**(6): 5165.

Lindholm, M., J. L. Hargraves, W. J. Ferguson, G. Reed. 2012. Professional language interpretation and inpatient length of stay and readmission rates. *Journal of General Internal Medicine* **27**(10): 1294–1299.

Othman, S. B., S. Hammadi, A. Quilliot. 2015. Multi-Objective Evolutionary for Multi-Skill Health Care Tasks Scheduling. *IFAC-PapersOnLine* **48**(3): 704–709.

Quan, K., J. Lynch. 2010. The high costs of language barriers in medical malpractice. *National Health Law Program* 1–24.

Rawal, S., J. Srighanthan, A. Vasantharoopan, H. Hu, G. Tomlinson, A. M. Cheung. 2019. Association Between Limited English Proficiency and Revisits and Readmissions After Hospitalization for Patients With Acute and Chronic Conditions in Toronto, Ontario, Canada. *JAMA* **322**(16): 1605–1607.

Shapiro, A., T. Homem-de-Mello. 2000. On the Rate of Convergence of Optimal Solutions of Monte Carlo Approximations of Stochastic Programs. *SIAM Journal on Optimization* **11**: 70–86.

Shapiro, A., T. Homem-de-Mello, J. Kim. 2002. Conditioning of convex piecewise linear stochastic programs. *Mathematical Programming* **94**: 1–19.

Simon, M. 2020, April 23. Language barriers can mean life or death in fight against coronavirus [Text]. TheHill. Available at https://thehill.com/changing-america/opinion/494284-isolated-with-no-family-members-for-non-english-speakers-with (accessed date July 25, 2020).

Ta, T. A., W. Chan, F. Bastin, P. L'Ecuyer. 2021. A simulation-based decomposition approach for two-stage staffing optimization in call centers under arrival rate uncertainty. *European Journal of Operational Research* **293**(3): 966–979.

Taffel, M. T., C. Huang, J. A. Karajgikar, K. Melamud, H. C. Zhang, A. B. Rosenkrantz. 2020. Retrospective analysis of the effect of limited english proficiency on abdominal MRI image quality. *Abdominal Radiology* 1–7.

Torresdey, P., J. Chen, H. P. Rodriguez. 2024. Patient Time Spent With Professional Medical Interpreters and the Care Experiences of Patients With Limited English Proficiency. *J Prim Care Community Health* **15**: 21501319241264168.

Tsai, C.-W., M.-C. Chiang. 2023. *Handbook of metaheuristic algorithms: from fundamental theories to advanced applications*. Elsevier.

Van den Bergh, J., J. Beliën, P. De Bruecker, E. Demeulemeester, L. De Boeck. 2013. Personnel scheduling: A literature review. *European Journal of Operational Research* **226**(3): 367–385.

Verweij, B., S. Ahmed, A. J. Kleywegt, G. Nemhauser, A. Shapiro. 2003. The sample average approximation method applied to stochastic routing problems: a computational study. *Computational Optimization and Applications* **24**: 289–333.

Wilson, C. C. 2013. Patient safety and healthcare quality: the case for language access. *International Journal of Health Policy and Management* **1**(4): 251.

Wilson, E., A. H. Chen, K. Grumbach, F. Wang, A. Fernandez. 2005. Effects of Limited English Proficiency and Physician Language on Health Care Comprehension. *J Gen Intern Med* **20**(9): 800–806.

Wilson-Stronks, A., E. Galvez. 2007. Exploring cultural and linguistic services in the nation's hospitals: A report of findings. *Oakbrook Terrace, IL: The Joint Commission*.